\documentclass[graybox]{svmult}

\usepackage{helvet}         % selects Helvetica as sans-serif font
\usepackage{courier}        % selects Courier as typewriter font
\usepackage{type1cm}        % activate if the above 3 fonts are
\usepackage{makeidx}         % allows index generation
\usepackage{graphicx}        % standard LaTeX graphics tool
\usepackage{multicol}        % used for the two-column index
\usepackage[bottom]{footmisc}% places footnotes at page bottom
\usepackage{amsmath,amssymb,calc}
\usepackage{subfig}
\usepackage{cleveref}
\usepackage{cite}

\newcommand{\be}{\begin{equation*}}
\newcommand{\ee}{\end{equation*}}
\newcommand{\ben}{\begin{equation}}
\newcommand{\een}{\end{equation}}
\newcommand{\beqa}{\begin{eqnarray*}}
\newcommand{\eeqa}{\end{eqnarray*}}
\newcommand{\beqan}{\begin{eqnarray}}
\newcommand{\eeqan}{\end{eqnarray}}
\newcommand{\nn}{\nonumber}

\newcommand{\pd}{\partial}

\def\dd{\mathrm{d}}
\newcommand{\eqdef}{\stackrel{\mathrm{def.}}{=}}

\def\id{\protect{{1 \kern-.28em {\rm l}}}}

\begin{document}

\title*{Consistency Condition for Slow-roll and Rapid-turn Inflation}
\titlerunning{Condition for Slow Roll $\&$ Rapid Turn}
% Use \titlerunning{Short Title} for an abbreviated version of
% your contribution title if the original one is too long
\author{Lilia Anguelova and Calin Lazaroiu}
\authorrunning{L. Anguelova and C. Lazaroiu}
% Use \authorrunning{Short Title} for an abbreviated version of
% your contribution title if the original one is too long
\institute{Lilia Anguelova \at Institute for Nuclear Research and Nuclear Energy, Bulgarian Academy of Sciences, Tsarigradsko Chaussee 72, Sofia 1784, Bulgaria,\\ \email{anguelova@inrne.bas.bg}
\and Calin Lazaroiu \at Department of Theoretical Physics, Horia Hulubei National Institute for Physics and Nuclear Engineering, Str. Reactorului no. 30, Bucharest-Magurele, Romania,\\ \email{lcalin@theory.nipne.ro}}
\maketitle

\abstract{We summarize our work on a consistency condition for inflation in two-field cosmological models, in the regime of rapid turn and third-order slow roll. To ensure a sustained inflationary period of this type, one needs to satisfy a certain relation between the scalar potential and the scalar field-space metric. We explain the derivation of this condition. Furthermore, we argue that, generically, the rapid-turn phase tends to be short-lived.}

\section{Introduction}
\label{sec:1}

Recent theoretical considerations \cite{GK,OPSV,AP,BPR} motivate the study of multifield inflationary models, which arise from the interaction of several scalar fields with gravity. These models are also significantly reacher phenomenologically, compared to single-field ones. This is due to the possibility of having background solutions, whose field-space trajectories deviate from geodesics. Such trajectories have a non-vanishing turn rate, which can lead to novel effects. For instance, a brief period of rapid turning can induce the generation of primordial black holes \cite{PSZ,FRPRW,LA,LA2}. Long-term rapid-turn inflationary models \cite{AB,SM,TB,BM,GSRPR,CRS,ACIPWW,APR} are also of great interest, for example with the aim of realizing slow roll inflation on steep potentials. The latter is desirable, in order to facilitate the embedding of the inflationary regime in a UV-complete description.

There is no guarantee, however, that the entire set of approximations, defining the rapid turn and slow roll regime, can be satisfied by actual background solutions for long enough, to produce full-fledged models of inflation. In fact, it was already noticed in \cite{ACPRZ} that it is difficult to find such rapid-turn inflationary solutions, albeit specifically in the context of supergravity. The conditions for compatibility between the relevant approximations and the equations of motion, in two-field cosmological models, were studied systematically in \cite{AL}. That work showed that there is a relation between the scalar potential and the scalar field-space metric, which needs to be satisfied in order to have a sustained rapid-turn and slow-roll inflationary period.

We review the derivation of the consistency condition, found in \cite{AL}, for long-lasting inflationary expansion in the regime of rapid turn and third-order slow roll. This condition is highly nontrivial and thus represents a severe constraint on the potential, for any given field-space metric. We also discuss the behavior of the so called characteristic angle, which determines the relation between two important bases in field space. One of those bases is defined in terms of the field-space trajectory of a background solution, while the other in terms of the potential and field space metric. We argue that the time evolution of the characteristic angle indicates that, generically, the rapid turn period can last for only a few e-folds.

\section{Equations of motion and inflationary parameters}
\label{sec:2}

We are interested in models of cosmological inflation obtained from coupling Einstein gravity to a set of real scalars. Let us begin by recalling the relevant action and introducing a number of important parameters that characterize the behavior of an inflationary solution. 

We will consider the minimally coupled case, which is described by the action: 
\ben
\label{Action_gen}
S = \int d^4x \sqrt{-\det g} \left[ \frac{R(g)}{2} - \frac{1}{2} G_{IJ} (\{\varphi^I\}) \pd_{\mu} \varphi^I \pd^{\mu} \varphi^J - V (\{ \varphi^I \}) \right] \,\,\, ,
\een
where $\varphi^I (x^{\mu})$ are the scalar fields, $g_{\mu \nu}$ is the spacetime
metric with $R(g)$ being its scalar curvature, $G_{IJ}$ is the metric on the manifold with coordinates $\{\varphi^I\}$ and the indices run as $\mu,\nu = 0,...,3$ and $I,J = 1,...,n$\,. We will take $n=2$ from now on, i.e. we will focus on the case of two scalars. Also, as usual, we will make the following Ansatze:
\ben
\label{metric_g}
\dd s^2_g = -\dd t^2 + a^2(t) \,\dd \vec{x}^2 \qquad , \qquad \varphi^I = \varphi^I (t) \quad ,
\een 
for the spacetime metric and scalars of the cosmological background. Here
$a(t)$ is the scale factor, which determines the Hubble parameter via the standard definition:
\ben
\label{Hp}
H (t) = \frac{\dot{a}}{a}
\een
with the dot denoting derivative with respect to $t$.

The action (\ref{Action_gen}) implies the following equations of motion for the background scalars:
\ben
\label{EoM_sc}
D_t \dot{\varphi}^I + 3 H \dot{\varphi}^I + G^{IJ} V_J = 0 \quad ,
\een
where $V_J \eqdef \pd_{\varphi^J} V$ and the derivative $D_t$ is defined as:
\ben
D_t A^I \eqdef \dot{\varphi}^J \,\nabla_J A^I = \dot{A}^I + \Gamma^I_{JK}(\varphi) \,\,\dot{\varphi}^J A^K
\een
for any vector $A^I$ on the scalar manifold (i.e., the manifold parameterized by the scalars $\varphi^I$). Here $\Gamma^I_{JK} (\varphi)$ are the Christoffel symbols of the metric $G_{IJ}$\,. In addition, it is convenient to write the Einstein equations as:
\ben
\label{EinstEqs}
G_{IJ} \dot{\varphi}^I \dot{\varphi}^J = - 2 \dot{H}  \qquad  ,  \qquad  3 H^2 + \dot{H} = V\quad .
\een

An important set of inflationary parameters is given by: 
\ben
\label{SR_par}
\varepsilon\eqdef-\frac{\dot{H}}{H^2} \qquad , \qquad \eta^I \eqdef - \frac{1}{H \dot{\sigma}} D_t \dot{\varphi}^I \quad .
\een
To understand the relevant physics, it is useful to decompose the vector $\eta^I$ in a suitable basis in field space. For that purpose, let us consider the field-space trajectory $\left( \varphi^1 (t), \varphi^2 (t) \right)$ of a background solution. Introducing the notation
\ben \label{sigma_dot}
\dot{\sigma} \equiv \sqrt{G_{IJ} \dot{\varphi}^I \dot{\varphi}^J}~~,
\een
one can define tangent and normal vectors to the trajectory by:
\ben
\label{N_def}
T^I = \frac{\dot{\varphi}^I}{\dot{\sigma}} \qquad {\rm and} \qquad N_I = (\det G)^{1/2} \epsilon_{IJ} T^J \,\,\, ,
\een
respectively. Notice that the quantity $\sigma$\,, determined by (\ref{sigma_dot}), is the proper length parameter of the curve $\left( \varphi^1 (t), \varphi^2 (t) \right)$\,. Also, the vectors $T^I$ and $N_I$\,, given in (\ref{N_def}), form an orthonormal basis in the scalar field space. With the help of this basis, we can define the turning rate of a trajectory: 
\ben
\label{Om_1}
\Omega \eqdef - N_I D_t T^I \,\,\, ,
\een
which characterizes its deviation from a geodesic. 

Now, expanding $\eta^I$ in the $(T,N)$ basis gives:
\ben
\eta^I = \eta_{\parallel} T^I + \eta_{\perp} N^I \,\,\, ,
\een
where
\ben
\label{eta_PP}
\eta_{\parallel} = - \frac{\ddot{\sigma}}{H \dot{\sigma}} \qquad {\rm and} \qquad \eta_{\perp} = \frac{\Omega}{H} \,\,\, .
\een
To obtain (\ref{eta_PP}), one has to use the definitions (\ref{N_def})-(\ref{Om_1}) and the projections of the field equations (\ref{EoM_sc}) along $T^I$ and $N_I$\,, respectively: 
\ben
\label{EoM_sigma}
\ddot{\sigma} + 3 H \dot{\sigma} + V_T = 0 \qquad , \qquad {\rm where} \,\,\, V_T \eqdef T^I V_I \,\,\, ,
\een
and
\ben
\label{EoM_N}
N_I D_t T^I = - \frac{V_N}{\dot{\sigma}} \qquad , \qquad {\rm where} \,\,\, V_N\eqdef N^I V_I \,\,\, .
\een
In particular, note that (\ref{Om_1}) and (\ref{EoM_N}) imply:
\ben
\label{Om_2}
\Omega = \frac{V_N}{\dot{\sigma}} \,\,\, .
\een

The quantities $\varepsilon$ and $\eta_{\parallel}$ introduced above coincide with the first and second slow roll parameters of single-field inflation with inflaton $\sigma (t)$\,, whereas the dimensionless turning rate $\eta_{\perp}$ is a distinguishing feature of the two-field case. In the following, we will also need the third slow roll parameter, defined by:
\ben \label{3rdOsr_par}
\xi\eqdef \frac{\dddot{\sigma}}{H^2\dot{\sigma}} \,\,\, ,
\een
as well as the second turning parameter:
\ben \label{nu_def}
\nu\eqdef \frac{\dot{\eta}_{\perp}}{H \eta_{\perp}} \,\,\, .
\een
Note that $\nu$ measures the rate of change of $\eta_{\perp}$ and, thus, is an important characteristic of the duration of a rapid-turn inflationary regime.

Finally, let us also define the conservative parameter:
\ben
\label{c_def}
c \eqdef \frac{H \dot{\sigma}}{\sqrt{G^{IJ} V_I V_J}} \,\,\, ,
\een
which is a measure for the relative strength of the friction term, compared to the gradient term in the scalar field equations. Note that $c > 0$ by definition.

\section{Consistency condition for slow roll and rapid turn}
\label{sec:3}

Taking derivatives of (\ref{EoM_sigma}) and (\ref{Om_2}), one can obtain the following exact relations (see \cite{HP,AGHPP,CCLBNZ}):
\beqan
\label{VTT_VTN}
\frac{V_{TT}}{3 H^2} &=& \frac{\eta_{\perp}^2}{3} + \varepsilon + \eta_{\parallel} - \frac{\xi}{3} \,\,\, , \nn \\
\frac{V_{TN}}{H^2} &=& \eta_{\perp} \left( 3 - \varepsilon - 2 \eta_{\parallel} + \nu \right) \,\,\, ,
\eeqan
where $V_{TT} = T^I T^J \nabla_I V_J$ and $V_{TN} = T^I N^J \nabla_I V_J$\,.
In \cite{AL}, it was shown that (\ref{VTT_VTN}) leads to a consistency condition for the existence of an inflationary regime with third-order slow roll, defined by:
\ben
\label{SR3}
\varepsilon \,, \,|\eta_{\parallel}| \,, \,|\xi| \ll 1 \,\,\, ,
\een
as well as sustained rapid turn, characterized by:
\ben
\label{SCT}
\eta_{\perp}^2 \gg 1 \quad , \quad |\nu| \ll 1 \,\,\, .
\een
Note that the condition on $|\eta_{\perp}|$ ensures that it is of order ${\cal O} (10)$ or larger, while the condition on $\nu$ ensures that the rapid-turn period is prolonged. 

The above mentioned consistency condition involves only the metric $G_{IJ}$ and the potential $V$\,. To explain its derivation, let us introduce a globally-defined orthonormal basis $(n,\tau)$ in field space by:
\ben
\label{ndef}
n^K = \frac{G^{KL} V_L}{\sqrt{G^{IJ} V_I V_J}} \qquad {\rm and} \qquad \tau_I = (\det G)^{1/2} \epsilon_{IJ} n^J \,\,\, .
\een
Note that $n$ is a unit vector along $\nabla V$ and that $(n,\tau)$ is positively oriented. The relation between this new basis and the trajectory-dependent one, defined in (\ref{N_def}), is given by:
\beqan
\label{BasesRel}
T &=& ~\,\cos \theta_{\varphi} \,\, n + \sin \theta_{\varphi} \,\, \tau \,\,\, , \nn \\
N &=& -\!\sin \theta_{\varphi} \,\, n + \cos \theta_{\varphi} \,\, \tau \,\,\, ,
\eeqan
where the time-dependent angle $\theta_{\varphi}\in (-\pi,\pi]$ determines the rotation from $(n,\tau)$ to the oriented basis $(T,N)$ at each point of the trajectory $(\varphi^1 (t),\varphi^2(t))$. 

For further use, let us now analyze what the approximations (\ref{SR3}) and (\ref{SCT}) imply in terms of the angle $\theta_{\varphi}$\,. From \eqref{eta_PP}-(\ref{EoM_sigma}) and (\ref{BasesRel}) one finds that $\eta_{\parallel} = 3 + \frac{\cos \theta_{\varphi}}{c}$\,, where $c$ is the parameter introduced in (\ref{c_def}). Hence the slow-roll condition $|\eta_{\parallel}| \ll 1$ implies:
\ben \label{c_cos_th}
c \approx - \cos \theta_{\varphi} / 3 \,\,\, .
\een
Similarly, from (\ref{eta_PP}), (\ref{Om_2}) and (\ref{BasesRel}), one obtains \cite{AL}:
\ben
\label{eta_perp_c}
\eta_{\perp} = - \frac{\sin \theta_{\varphi}}{c} \,\,\, .
\een
Using (\ref{c_cos_th}) in (\ref{eta_perp_c}) gives:
\ben
\label{eta_perp_tan}
\eta_{\perp}^2 \approx 9 \,\tan^2 \theta_{\varphi} \,\,\, .
\een
Therefore, to have rapid turning, during slow roll, we need $\tan^2 \theta_{\varphi} \gg 1$\,. The latter is equivalent with $\cos^2
\theta_{\varphi} \ll 1$\,, which implies $c^2 \ll 1$ due to (\ref{c_cos_th}).\footnote{Despite that, note that only $c < 1$ is needed, but not $c \ll 1$\,.}
In view of that, during slow roll the approximations (\ref{SCT}) are equivalent with
\ben
\label{SRRT}
c^2 \ll 1   \quad , \quad |\nu| \ll 1 \,\,\, . 
\een

With the above preparation, we are ready to turn to the derivation of the consistency condition for third order slow-roll and rapid-turn inflation. In the regime (\ref{SR3})-(\ref{SCT}), relations (\ref{VTT_VTN}) become:
\beqan
\label{VTN_c}
V_{TT} &\approx & \frac{V}{3\,c^2} - 3 V \,\,\, , \nn \\
V_{TN} &\approx & - \frac{s}{c} \sqrt{1-9c^2}\,V \,\,\, ,
\eeqan
where we used (\ref{EinstEqs}) and (\ref{c_cos_th})-(\ref{eta_perp_c}); more precisely, we used that (\ref{c_cos_th}) implies that $\sin \theta_{\varphi} \approx s (1 - 9 c^2)^{1/2}$ with $s = \pm 1$\,, while (\ref{EinstEqs}) gives $3 H^2 \approx V$ during slow roll. Now let us transform $V_{TT} = T^I T^J \nabla_I V_J$ and $V_{TN} = T^I N^J \nabla_I V_J$ to the $(n,\tau)$ basis. Using (\ref{BasesRel}), we have:
\beqan
\label{VTN-Vntau}
\hspace*{-0.4cm}V_{TT} &=& V_{nn} \cos^2 \theta_{\varphi} + 2 V_{n \tau} \sin \theta_{\varphi} \cos \theta_{\varphi} + V_{\tau \tau} \sin^2 \theta_{\varphi} \,\,\, , \nn \\
\hspace*{-0.4cm}V_{TN} &=& \left( V_{\tau \tau} - V_{nn} \right) \cos \theta_{\varphi} \sin \theta_{\varphi} + V_{n \tau} \left( \cos^2 \theta_{\varphi} - \sin^2 \theta_{\varphi} \right) \,\, ,
\eeqan
where $V_{nn} = n^I n^J \nabla_I V_J$ etc. Substituting (\ref{c_cos_th}) in (\ref{VTN-Vntau}) gives:
\beqan
\label{VTN-Vntau_c}
\hspace*{-0.4cm}V_{TT} &\approx & 9 c^2 V_{nn} - 6 s c \,\sqrt{1-9c^2} \,V_{n \tau} + (1-9c^2) V_{\tau \tau} \,\,\, , \nn \\
\hspace*{-0.4cm}V_{TN} &\approx & - 3 s c \,\sqrt{1-9c^2} \left( V_{\tau \tau} - V_{nn} \right) - (1-18c^2) V_{n \tau} \,\,\, .
\eeqan
Now, using (\ref{VTN-Vntau_c}) in (\ref{VTN_c}), we obtain:
\beqan
\label{VTN-Vntau_V_c}
\frac{V}{3\,c^2} - 3 V &\approx & 9 c^2 V_{nn} - 6 s c \,\sqrt{1-9c^2} \,V_{n \tau} + (1-9c^2) V_{\tau \tau} \,\,\, , \nn \\
\frac{s}{c} \sqrt{1-9c^2}\,V &\approx & 3 s c \,\sqrt{1-9c^2} \left( V_{\tau \tau} - V_{nn} \right) + (1-18c^2) V_{n \tau} \,\,\, .
\eeqan
From these two relations, we can solve for $c$ and then, using the result, derive a condition that involves only the potential and the components of its Hessian. 

More precisely, after some tedious manipulations and using (\ref{SRRT}), one can extract from the coupled system (\ref{VTN-Vntau_V_c}) the following relation \cite{AL}:
\ben \label{c_VntVnn}
V_{n \tau} \approx 3sc V_{nn} \,\,\, .
\een 
This implies, in particular, that if $V_{n \tau} \approx 0$ then $V_{nn} \approx 0$\,, and vice versa. When $V_{n\tau}, V_{nn} \neq 0$\,, one can solve (\ref{c_VntVnn}) for $c$\,. Substituting the result back in (\ref{VTN-Vntau_V_c}) leads to the following relation:
\ben
\label{VCond}
3 V V_{nn}^2 \approx V_{n \tau}^2 V_{\tau \tau} \,\,\, ,
\een
together with the inequality \cite{AL}:
\ben \label{ConsCond_Ineq}
V_{\tau \tau} |V_{nn}| \gg V_{n \tau}^2 \,\,\, .
\een 
Note that for $V_{n \tau}, V_{nn} \approx 0$ the relation (\ref{VCond}) is still satisfied (albeit trivially), although (\ref{ConsCond_Ineq}) is violated. For future use, notice also that (\ref{VCond}) implies: 
\ben \label{Vtt_pos}
V_{\tau \tau} > 0 \,\,\, ,
\een
due to $V > 0$ during inflation. 

Relation (\ref{VCond}) is a consistency condition for the inflationary regime of rapid turn with third order slow roll. Note that it relates the scalar potential $V$ to the field-space metric $G_{IJ}$\,, since the latter enters through the definitions $V_{n\tau} = n^I \tau^J \nabla_I V_J$ etc. Hence (\ref{VCond}) constrains severely the form of the potential for each choice of metric $G_{IJ}$\,, and vice versa. In the Appendix, we illustrate how this condition works in practice, by applying it to three prominent inflationary models.

\section{Evolution of the characteristic angle}
\label{sec:4}

The rather involved condition (\ref{VCond}) clearly indicates that, for a given metric $G_{IJ}$\,, only very special choices of $V$ are compatible with sustained slow roll and rapid turn inflation. Additional evidence to that effect is provided by the evolution with time of the characteristic angle $\theta_{\varphi}$\,. To investigate the latter, let us rewrite the equations of motion (\ref{EoM_sc}) in the basis (\ref{ndef}). That will allow us to derive an ODE that determines the function $\theta_{\varphi} (t)$\,.

We begin by expanding $\dot{\varphi}^I = \dot{\sigma} T^I$ as:
\ben \label{phiI_ntau}
\dot{\varphi}^I = v_n n^I + v_{\tau} \tau^I \,\,\, ,
\een
where due to (\ref{BasesRel}) one has:
\ben
\label{dtphi_ntau}
v_n \eqdef n_I \dot{\varphi}^I = \dot{\sigma} \cos \theta_{\varphi} \qquad {\rm and} \qquad v_{\tau} \eqdef \tau_I \dot{\varphi}^I = \dot{\sigma} \sin \theta_{\varphi} \,\,\, .
\een
Using (\ref{phiI_ntau}) in the first term of (\ref{EoM_sc}), one finds \cite{AL}:
\ben
\label{Dtphi_ntau}
D_t \dot{\varphi}^I = \left( \dot{v}_n - \lambda v_{\tau}^2 - \mu v_n v_{\tau} \right) n^I +
\left( \dot{v}_{\tau} + \mu v_n^2 + \lambda v_n v_{\tau} \right) \tau^I \,\,\, ,
\een
where:
\ben
\label{la_mu}
\lambda = - \frac{V_{\tau \tau}}{\sqrt{G^{IJ} V_I V_J}} \qquad , \qquad \mu = \frac{V_{n \tau}}{\sqrt{G^{IJ} V_I V_J}} \,\,\, .
\een
Due to (\ref{ndef}), the third term of (\ref{EoM_sc}) becomes:
\ben
\label{Vn}
G^{IJ} V_J \,= \,\sqrt{G^{KL}V_K V_L} \,\,\, n^I \,\,\, .
\een
Now, using (\ref{phiI_ntau}), (\ref{Dtphi_ntau}) and (\ref{Vn}), one finds for
the projections of (\ref{EoM_sc}) along the vectors $n$ and $\tau$\,:
\beqan
\label{EoMs_ntau}
\dot{v}_n - \lambda v_{\tau}^2 - \mu v_n v_{\tau} + 3H v_n + \sqrt{V_I V^I} &=& 0 \,\,\, , \nn \\
\dot{v}_{\tau} + \mu v_n^2 + \lambda v_n v_{\tau} + 3H v_{\tau} &=& 0 \,\,\, ,
\eeqan
respectively. Finally, substituting (\ref{dtphi_ntau}) in the second equation of (\ref{EoMs_ntau}) gives:
\ben
\label{th_d}
\dot{\theta}_{\varphi} \cos \theta_{\varphi} + (3 - \eta_{\parallel}) H \sin \theta_{\varphi} + \dot{\sigma} (\mu  \cos^2 \theta_{\varphi} + \lambda \sin \theta_{\varphi} \cos \theta_{\varphi} ) = 0 \,\,\, ,
\een
where we also used (\ref{eta_PP}). For convenience, one can rewrite (\ref{th_d}) as:
\ben
\label{th_d_E1}
\dot{\theta}_{\varphi} + (3 - \eta_{\parallel}) H \tan \theta_{\varphi} + \dot{\sigma} (\mu  \cos \theta_{\varphi} + \lambda \sin \theta_{\varphi} ) = 0 \,\,\, .
\een
Note that, in the slow roll approximation, substituting (\ref{dtphi_ntau}) in the first equation of (\ref{EoMs_ntau}) leads again to (\ref{th_d_E1}) \cite{AL}.

Let us now investigate the implications of (\ref{th_d_E1}) for the inflationary regime of interest, namely slow roll and rapid turn. To that effect, notice first that (\ref{c_def}) and (\ref{c_cos_th}) imply:
\ben \label{s_dot_cos_th}
\dot{\sigma} \approx - \frac{\cos \theta_{\varphi}}{3} \frac{\sqrt{V_I V^I}}{H} \,\,\, .
\een
Substituting (\ref{la_mu}), (\ref{s_dot_cos_th}) and $3 H^3 \approx V$ inside (\ref{th_d_E1}), we find that during slow roll it acquires the form:
\ben \label{Char_ang_Vnt_Vtt}
\dot{\theta}_{\varphi} + \sqrt{3 V} \tan \theta_{\varphi} - \frac{1}{\sqrt{3V}} \left( V_{n \tau} \cos^2 \theta_{\varphi} - V_{\tau \tau} \cos \theta_{\varphi} \sin \theta_{\varphi} \right) \approx 0 \,\,\, .
\een
Of course, to solve explicitly (\ref{Char_ang_Vnt_Vtt}) for $\theta_{\varphi} (t)$\,, one needs to consider specific potentials $V$ and metrics $G_{IJ}$\,. Nevertheless, we can make some generic inferences about the behavior of $\theta_{\varphi}$ by simplifying (\ref{Char_ang_Vnt_Vtt}) in the following manner. Recall that, according to the discussion below equation (\ref{eta_perp_tan}), rapid turning requires $\cos^2 \theta_{\varphi} \ll 1$\,, even though $\cos \theta_{\varphi}$ itself does not have to be small. Furthermore, in view of (\ref{ConsCond_Ineq}), it makes sense to assume that $V_{n \tau}$ is of order $V_{\tau \tau}$ or smaller. Hence, (\ref{Char_ang_Vnt_Vtt}) can be simplified to:
\ben \label{Char_ang_Vtt}
\dot{\theta}_{\varphi} + \sqrt{3 V} \tan \theta_{\varphi} + \frac{1}{\sqrt{3V}} V_{\tau \tau} \cos \theta_{\varphi} \sin \theta_{\varphi} \approx 0 \,\,\, .
\een

To gain insight in the solutions of (\ref{Char_ang_Vtt}), we will assume that (at least initially) $V \approx const$ along the inflationary trajectory\footnote{Due to (\ref{BasesRel}), this assumption means that $|\theta_{\varphi} (0)| \approx \frac{\pi}{2}$\,.}, which is reasonable during slow roll. For simplicity, we also take $V_{\tau \tau} \approx const$\,. Then the solution of (\ref{Char_ang_Vtt}) is determined by:
\ben \label{th_varph_cr_AB}
\cos \theta_{\varphi} \, \approx \, \pm \,\sqrt{\frac{\hat{C} e^{2\,(A+B)\,t} + A}{\hat{C} e^{2\,(A+B)\,t} - B}} \quad ,
\een
where $\hat{C} = const > 0$ and we have denoted:
\ben
A = \sqrt{3 V} \qquad , \qquad B = \frac{V_{\tau \tau}}{\sqrt{3 V}} \quad .
\een
Needless to say, (\ref{th_varph_cr_AB}) gives a crude estimate. Nevertheless, it leads to valuable intuition about the behavior of $\theta_{\varphi} (t)$\,. Namely, with time $\cos \theta_{\varphi}$ tends to $\pm 1$\,. Note that $\dot{\theta}_{\varphi} (t) > 0$ when $\theta_{\varphi} (0) = \frac{\pi}{2}$\,, whereas $\dot{\theta}_{\varphi} (t) < 0$ when $\theta_{\varphi} (0) = -\frac{\pi}{2}$\,; see \cite{AL}. Hence, in either
case $\theta_{\varphi}$ tends to $\pi \!\!\mod 2\pi$ with increasing $t$\,, implying
that the inflationary trajectory tends to align with $- \nabla V$\,. The characteristic timescale for that is $T = \frac{1}{A+B} =
\frac{\sqrt{3V}}{3V + V_{\tau \tau}}$\,. The above considerations suggest that, generically, the regime of rapid turning, during slow roll, is transient in nature. Of course, its precise duration will depend on the potential $V$ and field-space metric $G_{IJ}$\,, in each specific model.

It is possible, however, to obtain a crude model-independent estimate for the number of e-folds ${\cal N}$ per characteristic time $T$\,, in the slow-roll and rapid-turn inflationary regime under consideration. Namely, taking $V, V_{\tau \tau} \approx {\rm const}$ as above and using the slow-roll relation $3 H^2 \approx V$\,, one finds:
\ben
{\cal N} = \int_0^T H \,\dd t \,\approx \,\frac{V}{3 V + V_{\tau \tau}} \,\lesssim \,\frac{1}{3} \,\,\, ,
\een
where the last step is due to (\ref{Vtt_pos}). This suggests that, generically, the rapid turn period is brief, lasting for only one or at most a few e-folds. Thus, we have additional evidence in support of our earlier conclusion that sustained rapid turn, during slow roll, can occur only for special pairs $(V,G_{IJ})$ of potential $V$ and field-space metric $G_{IJ}$\,, but not generically.

\section{Concluding remarks}
\label{sec:5}

The consistency condition (\ref{VCond}) implies that, for a given metric $G_{IJ}$, only a particular form of the potential $V$ is compatible with a prolonged period of rapid-turn slow-roll inflation. Thus, the landscape of inflationary models of this type is much more limited than previously thought. It is, undoubtedly, very interesting to investigate what potentials are compatible, in this context, with classes of metrics $G_{IJ}$ that appear in phenomenologically-motivated string compactifications. Conversely, it is worth exploring what kinds of metrics $G_{IJ}$ are compatible with axionic, and other, potentials of interest.

Finally, it should be pointed out that, in the inflationary model-building literature, the slow-roll approximation is often discussed only in terms of the parameters $\varepsilon$ and $\eta_{\parallel}$\,. In other words, the third-order slow roll condition $\xi\ll 1$ is not taken into account. In view of our results, such models merit reconsideration, although one should keep in mind that current observations cannot place constraints on $\xi$\,. 

\begin{acknowledgement}
L.A. has received partial support from the Bulgarian NSF grant KP-06-N38/11. The work of C.L. was supported by grant PN 19060101/2019-2022.
\end{acknowledgement}

\section*{Appendix}
\addcontentsline{toc}{section}{Appendix}
To illuminate how the consistency condition (\ref{VCond}) is realized in practice, we consider three main examples of rapid turn inflation in the literature. One of them will turn out to have $V_{n \tau} = 0$\,, while for the other two we will show analytically that the approximate relation (\ref{VCond}) is well-satisfied within the relevant assumptions. 

\vspace{0.4cm}
\noindent
$\bullet$ \hspace*{0.03cm}{\bf Hyperinflation:}

\vspace{0.2cm}
\noindent
Let us begin with the example that is simplest technically, namely that of hyperinflation \cite{AB}. In this case, the field-space metric $G_{IJ}$ and the scalar potential $V$ are given by:
\ben \label{Hyp_infl_GV}
ds^2_G = d \phi^2 + L^2 \sinh^2 \!\left( \frac{\phi}{L} \right) d \theta^2 \quad, \quad V = V (\phi) \quad .
\een
The non-vanishing Christoffel symbols for this $G_{IJ}$ are:
\ben \label{Hyp_infl_Chr}
\Gamma^{\phi}_{\theta \theta} = - L \sinh \!\left( \frac{\phi}{L} \right) \cosh \!\left( \frac{\phi}{L} \right) \quad , \quad \Gamma^{\theta}_{\phi \theta} = \frac{1}{L} \coth \!\left( \frac{\phi}{L} \right) \quad .
\een
Substituting (\ref{Hyp_infl_GV}) in (\ref{ndef}), it is easy to obtain:
\ben \label{Hyp_infl_b}
n^{\phi} = \frac{V_{\phi}}{|V_{\phi}|} \quad , \quad n^{\theta} = 0 \quad , \quad \tau^{\phi} = 0 \quad , \quad \tau^{\theta} = - \frac{n^{\phi}}{L \sinh (\phi / L)} \quad .
\een
Using (\ref{Hyp_infl_Chr})-(\ref{Hyp_infl_b}) inside $V_{nn} = n^I n^J \nabla_I V_J$ etc., we find:
\ben
V_{nn} = V_{\phi \phi} \quad , \quad V_{n \tau} = 0 \quad , \quad V_{\tau \tau} = \frac{1}{L} \coth \!\left( \frac{\phi}{L} \right) \!V_{\phi} \quad .
\een

Since in this example $V_{n \tau}$ vanishes identically, while $V_{nn} \neq 0$\,, the consistency condition (\ref{VCond}) cannot be satisfied, strictly speaking. It would be interesting to investigate in the future which of the approximations in (\ref{SR3})-(\ref{SCT}) cannot be sustained in this case. Note however that, if
\ben \label{Hyp_infl_c}
3 V V_{\phi \phi} \ll 1 \quad ,
\een
then one can view (\ref{VCond}) as being well-satisfied numerically, up to a desired level of accuracy. Thus, the inequality (\ref{Hyp_infl_c}) is a necessary condition for a prolonged period of hyperinflation, in the rapid-turn and slow-roll regime defined by (\ref{SR3})-(\ref{SCT}). Notice that (\ref{Hyp_infl_c}) is somewhat different from the condition that was previously thought to be enough, namely $V_{nn} \ll V_{\tau \tau}$ (see \cite{TB}). 

\vspace{0.4cm}
\noindent
$\bullet$ \hspace*{0.03cm}{\bf Side-tracked inflation:}

\vspace{0.2cm}
\noindent
As a second example, we consider side-tracked inflation \cite{GSRPR}. In this case, the field-space metric $G_{IJ}$ and the scalar potential $V$ are given, respectively, by:
\ben \label{Side-tr_G}
ds^2_G = \left( 1 + \frac{2 \chi^2}{M^2} \right) d\phi^2 + d \chi^2 
\een
and
\ben \label{Side-tr_V}
V (\phi, \chi) = U (\phi) + \frac{m_h^2}{2} \chi^2 \quad .
\een
Also, the slow roll approximations $\varepsilon \ll 1$ and $|\eta_{\parallel}| \ll 1$ have the following forms, respectively \cite{GSRPR}:
\ben \label{Side-tr_SR}
m_h M |U'| \ll U^{3/2} \qquad {\rm and} \qquad m_h M U'' \ll |U'| U^{1/2} \quad ,
\een
where we have denoted $U' \equiv dU/d\phi$\,. Note that the Ricci scalar of the metric (\ref{Side-tr_G}) is $R_{G} = - \frac{4M^2}{(M^2+2\chi^2)^2}$\,. Hence, for small $\chi$ (as in side-tracked inflation), the parameter $M$ sets the scale of curvature of the field-space metric. The rapid-turn inflationary trajectory is given by \cite{GSRPR}:
\ben \label{Side-tr_traj}
\chi^2 = \frac{M^2}{2} \left( \sqrt{\frac{2}{3}} \frac{|U'|}{m_h M \sqrt{U}} - 1 \right) \quad .
\een
Note that, due to the first inequality in (\ref{Side-tr_SR}), the potential $V$ in (\ref{Side-tr_V}) is dominated by $U(\phi)$ along the inflationary trajectory (\ref{Side-tr_traj}). Another important feature of side-tracked inflation is that, unlike the light field $\phi$\,, the field $\chi$ is heavy. Namely, the mass of $\chi$ satisfies $m_h \gg H$\,, where $H$ is the Hubble parameter. Since during slow roll $3 H^2 \approx V$ (in principle) and during side-tracked inflation $V \approx U$ (as explained above), the inequality $m_h \gg H$ is equivalent with:
\ben \label{mh2_gg_U}
m_h^2 \gg U \quad .
\een
This relation will be very important in the following.

We now turn to the computations relevant for verifying the consistency condition (\ref{VCond}). The non-vanishing Christoffel symbols of the metric (\ref{Side-tr_G}) are:
\ben \label{Side-tr_Chr}
\Gamma^{\phi}_{\phi \chi} = \frac{2 \chi}{M^2 + 2 \chi^2} \quad , \quad \Gamma^{\chi}_{\phi \phi} = - \frac{2 \chi}{M^2} \quad .
\een
For convenience, let us introduce the notation:
\ben \label{nt_hat_def}
n^I \equiv \frac{\hat{n}^I}{\sqrt{N}} \quad {\rm and} \quad \tau^I \equiv \frac{\hat{\tau}^I}{\sqrt{N}} \quad , \quad {\rm where} \quad N \equiv G^{IJ} V_I V_J \quad .
\een
Then, using (\ref{Side-tr_G})-(\ref{Side-tr_V}) inside (\ref{ndef}), we obtain the following for the quantities defined in (\ref{nt_hat_def}):
\ben \label{Side-tr_b}
\hat{n}^{\phi} = \frac{V_{\phi}}{1+\frac{2\chi^2}{M^2}} \,\,\, , \,\,\, \hat{n}^{\chi} = V_{\chi} \,\,\, , \,\,\, \hat{\tau}^{\phi} = \frac{V_{\chi}}{\sqrt{1+\frac{2 \chi^2}{M^2}}} \,\,\, , \,\,\, \hat{\tau}^{\chi} = - \frac{V_{\phi}}{\sqrt{1+\frac{2 \chi^2}{M^2}}}
\een
and
\ben \label{Side-tr_N}
N = \frac{U'^2}{1+\frac{2 \chi^2}{M^2}} + m_h^4 \chi^2 \quad .
\een
It is also convenient to introduce $\hat{V}_{nn} \equiv \hat{n}^I \hat{n}^J \nabla_I V_J$ and similarly for $\hat{V}_{n \tau}$ and $\hat{V}_{\tau \tau}$\,. Using (\ref{Side-tr_V}), (\ref{Side-tr_Chr}) and (\ref{Side-tr_b}), we compute: 
\beqan
\hat{V}_{nn} &=& \frac{U'^2 \!\left( U'' - \frac{2 m_h^2 \chi^2}{M^2} \right)}{\left( 1 + \frac{2 \chi^2}{M^2} \right)^{\!2}} + m_h^6 \chi^2 \quad , \label{V_h_nn}\\
\hat{V}_{n \tau} &=& \frac{2 \,\chi \,U'^3}{M^2 \!\left( 1 + \frac{2 \chi^2}{M^2} \right)^{\!5/2}} + \frac{m_h^2 \,\chi \,U' \,U''}{\left( 1 + \frac{2 \chi^2}{M^2} \right)^{\!3/2}} - \frac{m_h^4 \,\chi \,U'}{\left( 1 + \frac{2 \chi^2}{M^2} \right)^{\!1/2}} \quad , \\
\hat{V}_{\tau \tau} &=& \frac{m_h^2 \,U'^2 \!\left( 1 + \frac{6 \chi^2}{M^2} \right)}{\left( 1 + \frac{2 \chi^2}{M^2} \right)^2} + \frac{m_h^4 \chi^2 \left( U'' + \frac{2 m_h^2 \chi^2}{M^2} \right)}{\left( 1 + \frac{2 \chi^2}{M^2} \right)} \label{V_h_tt} \quad .
\eeqan
Notice that $\hat{V}_{nn} = N V_{nn}$\,, $\hat{V}_{n \tau} = N V_{n \tau}$ and $\hat{V}_{\tau \tau} = N V_{\tau \tau}$\,. Hence the consistency condition (\ref{VCond}) can be written equivalently as:
\ben \label{Side-tr_CC}
3 V \hat{V}_{nn}^2 N \approx \hat{V}_{n \tau}^2 \hat{V}_{\tau \tau} \quad .
\een
Note also that $\chi$ enters both sides of (\ref{Side-tr_CC}) only via powers of $\chi^2$, as is clear from (\ref{Side-tr_N})-(\ref{V_h_tt}). 

Now let us compute the expressions on the two sides of the consistency relation (\ref{Side-tr_CC}). Using (\ref{Side-tr_N})-(\ref{V_h_tt}) and substituting the inflationary trajectory (\ref{Side-tr_traj}), as well as $V \approx U$, we obtain for the left-hand side:
\beqan \label{LHS_side-tr}
3 V \hat{V}_{nn}^2 N &\approx& \frac{\sqrt{6} M^3 m_h^7}{12} \left( \frac{m_h^6}{U^3} - 3 \frac{m_h^4}{U^2} - 9 \frac{m_h^2}{U} + 27 \right) U^{5/2} |U'|^3 \nn \\
&-& \frac{3}{4} M^4 m_h^8 \left( \frac{m_h^6}{U^3} - 4 \frac{m_h^4}{U^2} - 3 \frac{m_h^2}{U} + 18 \right) U^3 U'^2 \nn \\
&+& \frac{3}{8} \sqrt{6} M^5 m_h^9 \left( \frac{m_h^6}{U^3} - 5 \frac{m_h^4}{U^2} +3 \frac{m_h^2}{U} + 9 \right) U^{7/2} |U'| \nn \\
&-& \frac{3}{8} M^6 m_h^{12} \left( \frac{m_h^4}{U^2} - 6 \frac{m_h^2}{U} + 9 \right) U^3 \quad ,
\eeqan
where we have neglected several terms containing $U''$ due to the condition $m_h^2 \gg U''$ (which means that $\phi$ is light, compared to the heavy field $\chi$). To be more precise, $U''$ is negligible in (\ref{LHS_side-tr}), because it enters that expression only in the combination $(U'' + m_h^2)$\,.

Similarly, using (\ref{V_h_nn})-(\ref{V_h_tt}) and substituting (\ref{Side-tr_traj}), we find for the right-hand side of (\ref{Side-tr_CC}):
\beqan \label{RHS_side-tr}
\hat{V}_{n \tau}^2 \hat{V}_{\tau \tau} &\approx& \frac{\sqrt{6} M^3 m_h^7}{12} \left(  \frac{m_h^6}{U^3} + 3 \frac{m_h^4}{U^2} - 45 \frac{m_h^2}{U} + 81 \right) U^{5/2} |U'|^3 \nn \\
&-& \frac{3}{4} M^4 m_h^8 \left( \frac{m_h^6}{U^3} - \frac{m_h^4}{U^2} - 63 \frac{m_h^2}{U} + 135 \right) U^3 U'^2 \nn \\
&+& \frac{3}{8} \sqrt{6} M^5 m_h^9 \left( \frac{m_h^6}{U^3} - 4 \frac{m_h^4}{U^2} -3 \frac{m_h^2}{U} + 18 \right) U^{7/2} |U'| \nn \\
&-& \frac{3}{8} M^6 m_h^{12} \left( \frac{m_h^4}{U^2} - 6 \frac{m_h^2}{U} + 9 \right) U^3 \quad ,
\eeqan
where again we have neglected a number of terms containing $U''$, but now due to the second inequality in (\ref{Side-tr_SR}). Namely, the $U''$ terms in (\ref{RHS_side-tr}) are negligible during slow roll, because they enter that expression only in the combination $(c_1 \,m_h M U'' + c_2 \, |U'| \sqrt{U})$ with $c_{1,2}$ being numerical constants.

Let us now compare (\ref{LHS_side-tr}) and (\ref{RHS_side-tr}). It is easy to notice that in the limit (\ref{mh2_gg_U}) their leading terms coincide. More precisely, for $\frac{m_h^2}{U} \gg 1$\,, we obtain:
\beqan
3 V \hat{V}_{nn}^2 N \approx \hat{V}_{n \tau}^2 \hat{V}_{\tau \tau} &\approx& \frac{\sqrt{6} M^3 m_h^{13}}{12} \frac{|U'|^3}{\sqrt{U}} - \frac{3}{4} M^4 m_h^{14} U'^2 \nn \\
&+& \frac{3}{8} \sqrt{6} M^5 m_h^{15} \sqrt{U} |U'| - \frac{3}{8} M^6 m_h^{16} U \quad .
\eeqan
From (\ref{LHS_side-tr})-(\ref{RHS_side-tr}) we can also see explicitly how decreasing the ratio $m_h^2/U$ leads to an increasing violation of the consistency condition. And conversely, increasing $m_h^2/U$ improves the numerical agreement between the two sides of (\ref{Side-tr_CC}). This underscores the importance of $\chi$ being heavy, in order to realize side-tracked inflation. It also suggests that larger ratios $m_h^2/U$ could lead to longer periods of, or larger turning rates during, such an inflationary expansion. Although, one should keep in mind that the consistency condition is only necessary, but may not be sufficient, for the existence of a sustained slow-roll and rapid-turn regime.  

\vspace{0.4cm}
\noindent
$\bullet$ \hspace*{0.03cm}{\bf Angular inflation:}

\vspace{0.2cm}
\noindent
The third example we consider is angular inflation \cite{CRS}. In that case, the field-space metric and potential are:
\ben \label{Ang_G_V}
ds^2_G = \frac{6 \alpha (d\phi^2 + d\chi^2)}{\left( 1 - \phi^2 - \chi^2 \right)^2} \quad , \quad V (\phi, \chi ) = \frac{\alpha}{6} \left( m_{\phi}^2 \phi^2 + m_{\chi}^2 \chi^2 \right) \quad .
\een
It is convenient to perform a change of variables in field space: 
\ben
\phi = r \cos \theta \quad , \quad \chi = r \sin \theta \quad ,
\een
as well as to introduce the parameter:
\ben
R \equiv \frac{m_{\chi}^2}{m_{\phi}^2} \quad .
\een
Then, the metric and potential in (\ref{Ang_G_V}) become: 
\ben \label{Ang_rth_GV}
ds^2_G = 6 \alpha \frac{(dr^2 + r^2 d\theta^2)}{\left( 1 - r^2 \right)^2} \quad , \quad V = \frac{\alpha}{6} \,m_{\phi}^2 \,r^2 \left( \cos^2 \!\theta + R \sin^2 \!\theta \right) \quad .
\een
Angular inflation is obtained within the parameter ranges:
\ben \label{Ang_par_constr}
\alpha \ll 1 \quad {\rm and} \quad R \gtrsim 10 \quad.
\een
The field-space trajectory of the inflationary solution is given by \cite{CRS}:
\ben \label{Ang_traj}
r^2 = 1 - \frac{9}{2} \frac{\left( \cot \theta + R \tan \theta \right)^2}{(R-1)^2} \quad .
\een

Computing the Christoffel symbols for the metric in (\ref{Ang_rth_GV}), we find the following non-vanishing components:
\ben \label{Ang_Chr}
\Gamma^r_{rr} = \frac{2\,r}{1-r^2} \quad, \quad \Gamma^r_{\theta \theta} = - \frac{r \left( 1+r^2 \right)}{1-r^2} \quad , \quad \Gamma^{\theta}_{r \theta} = \frac{\left( 1+r^2 \right)}{r\left( 1-r^2 \right)} \quad .
\een
Using the same notation as in (\ref{nt_hat_def}), we also obtain from (\ref{Ang_rth_GV}) that:
\beqan
\hat{n}^r = \frac{\left( 1-r^2 \right)^2 V_r}{6 \,\alpha} \quad &,& \quad \hat{n}^{\theta} = \frac{\left( 1-r^2 \right)^2 V_{\theta}}{6 \,\alpha \,r^2} \,\,\, , \nn \\
\hat{\tau}^r = \frac{\left( 1-r^2 \right)^2 V_{\theta}}{6 \,\alpha r} \quad &,& \quad \hat{\tau}^{\theta} = - \frac{\left( 1-r^2 \right)^2 V_r}{6 \,\alpha \,r}
\eeqan
and
\ben \label{Ang_N}
N =\frac{\left( 1-r^2 \right)^2}{6 \alpha} \left( V_r^2 + \frac{V_{\theta}^2}{r^2} \right) \quad .
\een
Now, using (\ref{Ang_Chr})-(\ref{Ang_N}) and substituting (\ref{Ang_traj}), we can compute again the full expressions on the left and right sides of the consistency condition, just like in the side-tracked example above. And again, we find complicated expressions with many terms. To illustrate the good approximate agreement, let us consider only the leading terms for small $\alpha$ and large $R$\,, in line with (\ref{Ang_par_constr}). To leading order in $\alpha \ll 1$ and large $R$\,, we find that both sides of (\ref{VCond}) give:
\ben
3 V V_{nn}^2 \approx V_{n \tau}^2 V_{\tau \tau} \approx \frac{1}{8} \,R^3 \,\alpha^{3} \,m_{\phi}^6 \,\sin^6 \!\theta \,\tan^4 \!\theta \quad ,
\een
in perfect agreement with each other. Of course, each side contains multiple subleading terms, with higher powers of $\alpha$ and/or lower powers of $R$. So decreasing $R$ or increasing $\alpha$ will lead to an increasing violation of the consistency relation.

Undoubtedly, further study of the consequences of the consistency condition for slow-roll and rapid-turn inflation, either regarding the above examples or more generally, is of great interest. We hope to report progress on this topic in the future.

\end{document}